\documentclass{iopart}
\usepackage{graphicx}
\usepackage{iopams}
 
 \def\bc{\begin{center}}          \def\ec{\end{center}}

\begin{document}
 \title[Generation of powerful terahertz emission]{Generation of powerful terahertz emission in a beam-driven strong plasma turbulence}
 \author{A V Arzhannikov, I V Timofeev}
 \address{Budker Institute of Nuclear Physics SB RAS, 630090, Novosibirsk, Russia \\
 Novosibirsk State University, 630090, Novosibirsk, Russia}
\ead{timofeev@ngs.ru}
 \begin{abstract}

Generation of terahertz electromagnetic radiation due to
coalescence of upper-hybrid waves in the long-wavelength region of
strong plasma turbulence driven by a high-current relativistic
electron beam in a magnetized plasma is investigated. The width of
frequency spectrum as well as angular characteristics of this
radiation for various values of plasma density and turbulence
energy are calculated using the simple theoretical model
adequately describing beam-plasma experiments at mirror traps. It
is shown that the power density of electromagnetic emission at the
second harmonic of plasma frequency in the terahertz range for
these laboratory experiments can reach the level of 1
$\mbox{MW/cm}^3$ with 1\% conversion efficiency of beam energy
losses to electromagnetic emission.

 \end{abstract}

\sloppy
\section{Introduction}
It is well known that a turbulent plasma is a source of
electromagnetic emission at the fundamental plasma frequency
$\omega_p$ and its second harmonic. These emission mechanisms have
been found to play an important role in different space phenomena
such as type III solar radio bursts \cite{gur,kru,rob,mel,li2} and
radiations of planet's magnetospheres \cite{gur2,cai}. Plasma
emissions at $\omega_p$ and $2\omega_p$ have been also observed in
laboratory beam-plasma experiments \cite{ben,hop,ben2,arz}. In
contrast to the problem of type III radio bursts, in which the
density of wave energy $W$ is saturated at weakly turbulent levels
$W/nT\sim 10^{-5}$ ($n$ is the plasma density, $T$ is the
temperature of plasma electrons), we focus our attention on the
regime of strong plasma turbulence $W/n T\sim 10^{-2}- 10^{-1}$,
which is more appropriate for laboratory experiments with powerful
electron beams \cite{vya,vya1,arz1}.

Beam-plasma experiments at mirror traps \cite{arz1,arz2} have
demonstrated that the relativistic electron beam with the typical
energy 1 MeV and current density 1 $\mbox{kA/cm}^2$ injected into
the plasma with $n=2\cdot 10^{14}\, \mbox{cm}^{-3}$ loses about
30\% of its energy over the length of 1 m. It means that the
averaged power density pumping by the beam to the plasma
turbulence can be estimated as $\sim 10\, \mbox{MW/cm}^3$. The
rate of beam energy losses, however, is not uniform over this
length. The estimate for the peak pumping power can be derived
assuming that a significant part of beam energy goes to excitation
of the large amplitude coherent wave-packet at the entrance to the
plasma column. Since the beam is decelerated over the packet
length $l\sim v_b/\Gamma$ by $\Delta v\sim v_b \Gamma/\omega_p$
\cite{tim}, where $\Gamma$ is the growth-rate of the two-stream
instability and $v_b$ is the beam velocity, the maximum power of
beam energy losses reaches the value $\sim 100\, \mbox{MW/cm}^3$.
Thus, if only 1\% of pumping power is able to convert to
electromagnetic radiation, the maximum power density of plasma
emission produced in beam-plasma experiments at mirror traps can
reach the range of 1 $\mbox{MW/cm}^3$. The same level of
conversion has been observed recently in laboratory experiments
\cite{arz} with the low plasma density $n=2\cdot 10^{14}\,
\mbox{cm}^{-3}$, that is why it allows us to suppose that, in a
denser plasma, a beam excited turbulence can be also considered as
an efficient source of powerful electromagnetic radiation. The aim
of this paper is to calculate the conversion efficiency as well as
absolute values of emission power for the case when second
harmonic plasma emission falls in the terahertz frequency range.

Our calculations of the spectral power of electromagnetic emission
produced in a turbulent magnetized plasma are based on the
theoretical model \cite{tim1}, in which it is assumed that
electromagnetic waves are generated predominantly in the source
region of strong plasma turbulence due to coalescence of
upper-hybrid waves. Since theoretical predictions have been found to
agree with recent experimental results \cite{arz} obtained at the
GOL-3 multimirror trap in the low density regime, we can use this
model to study spectral and angular characteristics of
electromagnetic emission for the whole ranges of plasma density and
turbulence energy, which can be achieved in our beam-plasma
experiments.

Thus, in Section 2, we formulate the main ideas of theoretical model
that takes into account not only spontaneous generation of
electromagnetic waves in coalescence processes, but also
contributions of induced inverse processes. In Section 3, we present
calculations of second harmonic emission power in the terahertz
frequency range and study whether this emission can escape from the
plasma. Our main results are summarized in concluding Section 4.

\section{Theoretical model}

To calculate the emission power from turbulent magnetized plasma we
use the model of strong plasma turbulence proposed in \cite{tim1}.
According to this model, we assume that most of wave energy is
concentrated in upper-hybrid modes, which occupy  the
long-wavelength region of turbulent spectrum with wavenumbers $k
<\sqrt{W/ n T}/ r_D $ ($r_D$ is the Debye length). The saturation
level of this energy is determined by the balance between the
constant power $P_b$ pumping to the turbulence by the beam and the
power dissipated  due to the wave collapse. It results in the
following relation between the pumping power and the turbulence
energy:
\begin{equation}\label{eq1}
    P_b\sim \omega_p n T
    \left(\frac{m_e}{m_i}\right)^{1/2}\left(\frac{W}{nT}\right)^2,
\end{equation}
where $m_{i}$ is the ion mass  and $m_{e}$ is the electron mass.

We calculate the electromagnetic emission power $P$ and the
conversion efficiency $\epsilon=P/P_b$ assuming that electromagnetic
waves are generated due to coalescence of long-wavelength
upper-hybrid waves. Since in the source region of turbulent spectrum
these waves are not trapped in collapsing caverns, nonlinear
interaction between electromagnetic and upper-hybrid modes can be
treated in the framework of weak turbulence. In this case,
generation of electromagnetic radiation is described by the equation
\begin{equation}\label{eq2}
    \frac{\partial W_k^t}{\partial t}=P_k-2 \gamma_k W_k^t,
\end{equation}
where $P_k$ is the power of spontaneous emission produced in the
coalescence process $\ell + \ell\rightarrow t$ and $\gamma_k$ is the
nonlinear dissipation rate due to the decay $t\rightarrow \ell +
\ell$. In dimensionless units $\omega_p t$, $\omega/\omega_p$, $x
\omega_p/c$, $kc/\omega_p$,  for time, frequency, position and
wavenumber, respectively, the spectral wave energy is normalized by
the condition
\begin{equation}\label{eq1a}
 \int W_k^{\sigma} d^3 k=\frac{W^{\sigma}}{n m_e c^2},
\end{equation}
where $c$ is the speed of light. In the cold plasma limit,
dimensionless values of $P_k$ and $\gamma_k$ take the forms
\begin{eqnarray}\label{eq3}
\fl P_k= \frac{2 \pi}{\omega_k^t (\partial \Lambda /\partial
\omega)_{\omega_k^t}} \int \frac{W_{k_1}^{\ell}
 W_{k_2}^{\ell} \left|G^{t \ell\ell}_{k, k_1, k_2}\right|^2 \Delta_{k, k_1, k_2}}{\omega_{k_1}^{\ell} (\partial
\Lambda /\partial \omega)_{\omega_{k_1}^{\ell}} \omega_{k_2}^{\ell}
(\partial \Lambda /\partial \omega)_{\omega_{k_2}^{\ell}}}\nonumber \\
\times\delta({\bf k}-{\bf k}_1-{\bf k}_2) d^3 k_1 d^3 k_2,
\end{eqnarray}
\begin{eqnarray}\label{eq1b}
\fl \gamma_k=\frac{1}{\omega_k^t (\partial \Lambda /\partial
\omega)_{\omega_k^t}} \int \frac{
 W_{k_2}^{\ell} \delta({\bf k}-{\bf k}_1-{\bf k}_2) }{\omega_{k_1}^{\ell} (\partial
\Lambda /\partial \omega)_{\omega_{k_1}^{\ell}} \omega_{k_2}^{\ell}
(\partial \Lambda /\partial \omega)_{\omega_{k_2}^{\ell}}} \nonumber \\
\times\left[\frac{-i G_{k,k_1,k_2}^{t \ell \ell}G_{k_1,-k_2,k}^{\ell
\ell t}}{\omega_{k_1}^{\ell}+\omega_{k_2}^{\ell}-\omega_{k}^{t}-i
\nu}+\mbox{c. c.} \right]
 d^3 k_1 d^3 k_2,
\end{eqnarray}
where we introduce the following notations:
\begin{equation}\label{eq4}
 \fl   \Lambda^{\sigma} ({\bf
k},\omega)= |{\bf k}\cdot {\bf e}^{\sigma}_k|^2 -k^2 +\omega^2
\left({\bf e}^{\ast \sigma}_k {\bf\hat{\varepsilon}}_{k}^{\sigma}
{\bf e}^{\sigma}_k\right),
\end{equation}
\begin{eqnarray}\label{eq5}
\fl  G^{\sigma \sigma^{\prime} \sigma^{\prime \prime}}_{k, k_1, k_2}
= \frac{\omega_{+}}{\omega_{k_1}^{\sigma^{\prime}}} \left({\bf
e}_{k}^{\ast \sigma} \widehat{T}_{k_2}^{\sigma^{\prime\prime}} {\bf
e}_{k_2}^{\sigma^{\prime\prime}}\right) \left({\bf k}_1
\widehat{T}_{k_1}^{\sigma^{\prime}} {\bf
e}_{k_1}^{\sigma^{\prime}}\right) \nonumber \\
+\frac{\omega_{+}}{\omega_{k_2}^{\sigma^{\prime\prime}}} \left({\bf
e}_{k}^{\ast \sigma} \widehat{T}_{k_1}^{\sigma^{\prime}} {\bf
e}_{k_1}^{\sigma^{\prime}}\right) \left({\bf k}_2
\widehat{T}_{k_2}^{\sigma^{\prime\prime}} {\bf
e}_{k_2}^{\sigma^{\prime\prime}}\right)  + {\bf e}_{k}^{\ast \sigma}
\widehat{T}^{+} {\bf g},
\end{eqnarray}
\begin{eqnarray}\label{eq6}
\fl    {\bf g}= \left({\bf k}_2 \widehat{T}_{k_1}^{\sigma^{\prime}}
{\bf e}_{k_1}^{\sigma^{\prime}}\right) \left[
\widehat{T}_{k_2}^{\sigma^{\prime\prime}} \cdot {\bf
e}_{k_2}^{\sigma^{\prime\prime}}-\left(1-\frac{\Omega^2}{(\omega_{k_2}^{\sigma^{\prime\prime}})^2}\right)
{\bf e}_{k_2}^{\sigma^{\prime\prime}}\right] \nonumber \\ +{\bf
k}_{2} \left({\bf e}_{k_2}^{\sigma^{\prime\prime}}
\widehat{T}_{k_1}^{\sigma^{\prime}} {\bf
e}_{k_1}^{\sigma^{\prime}}\right)+(k_1,
\sigma^{\prime}\rightleftarrows k_2, \sigma^{\prime\prime}),
\end{eqnarray}
\begin{equation}\label{eq7}
  \fl  \widehat{T}_k^{\sigma}=\left(\omega_k^{\sigma}\right)^2
    \left(\hat{I}-\hat{\varepsilon}_k^{\sigma}\right), \qquad \omega_{+}
=\omega^{\sigma^{\prime}}_{k_1}+\omega^{\sigma^{\prime\prime}}_{k_2}.
\end{equation}
Here, eigenfrequencies $\omega_k^{\sigma}$ and eigenvectors ${\bf
e}_k^{\sigma}$ of linear plasma modes are determined by the standard
dispersion equation with the dielectric tensor
$\hat{\varepsilon}_k^{\sigma}$ ($\hat{I}$ is the unit matrix). In
contrast to the similar calculations of second harmonic emission
\cite{wil,kuz} based on the standard weak turbulence theory, we take
into account model damping of two-time correlation functions, which
is used to describe the effect of finite life-time of upper-hybrid
plasmons due to their scattering off density fluctuations with the
typical frequency $\nu=\omega_p W^{\ell}/(nT)$. It results in
correlation broadening of the resonance
$\omega_k^t-\omega_{k_1}^{\ell}-\omega_{k_2}^{\ell}=0$, which is
described in (\ref{eq3}) by the function
\begin{equation}\label{eq8}
\Delta_{k, k_1, k_2}=\frac{2 \nu /\pi}{\left(\omega_k^t-
\omega_{k_1}^{\ell} -\omega_{k_2}^{\ell}\right)^2+4 \nu^2}.
\end{equation}
This function shows that the width of frequency spectrum of second
harmonic electromagnetic emission depends essentially on nonlinear
effects as well as on thermal and magnetic corrections to the linear
dispersion of upper-hybrid modes. In order to minimize the frequency
width of radiation, we should consider the case, when the ratio of
the electron cyclotron frequency to the plasma frequency becomes low
$\Omega=\omega_c/\omega_p=0.2$ resulting in comparable contributions
of magnetic and thermal effects for the typical plasma temperature
$T=1$ keV in mirror traps. As to account for the effect of finite
temperature, we modify the eigenfrequencies and eigenvectors of
linear plasma modes, but we neglect modifications in the nonlinear
current $G^{t \ell\ell}_{k, k_1, k_2}$, which is much less sensitive
to thermal corrections than $\Delta_{k, k_1, k_2}$. Thus, to
calculate $\omega_k^{\sigma}$, we use the fluid approximation and
solve the linear dispersion relation with the dielectric tensor:
\begin{eqnarray}
\varepsilon_{xx} = 1-A \left(1-\frac{k_{\parallel}^2
V_T^2}{\omega^2}\right), \\
\varepsilon_{xy} = -\varepsilon_{xy}=i \frac{\Omega}{\omega} A
\left(1-\frac{k_{\parallel}^2 V_T^2}{\omega^2}\right), \\
\varepsilon_{yy} = 1-A \left(1-\frac{k^2 V_T^2}{\omega^2}\right),\\
\varepsilon_{xz} = \varepsilon_{zx}= -A \frac{k_{\parallel}
k_{\perp} V_T^2}{\omega^2}, \\
\varepsilon_{yz} = -\varepsilon_{zy}=-i \frac{\Omega}{\omega} A
\frac{k_{\parallel} k_{\perp} V_T^2}{\omega^2}, \\
\varepsilon_{zz} = 1-A \left(1-\frac{k_{\perp}^2
V_T^2+\Omega^2}{\omega^2}\right), \\
A = \left(\omega^2-\Omega^2-k^2 V_T^2+\frac{\Omega^2}{\omega^2}
k_{\parallel}^2 V_T^2\right)^{-1}, \nonumber
\end{eqnarray}
where $V_T^2=3T/(m_e c^2)$, magnetic field is directed along
$z$-axis and ${\bf k}=(k_{\perp}, 0, k_{\parallel})$  is the wave
vector with the length $k=(k_{\perp}^2+k_{\parallel}^2)^{1/2}$.

If the path lengths of spontaneously generated electromagnetic waves
are larger than the typical size of confined plasma
$l_k=v_g/\gamma_k>L$ ($v_g$ is the group velocity of electromagnetic
wave), decay processes $t\rightarrow \ell + \ell$ do not play a
role, and the second term in (\ref{eq2}) can be omitted. Thus, in
the case of azimuthally symmetric turbulence, the spectral emission
power in units of $nm_e c^2$ is given by the integral
\begin{equation}\label{eq11}
    \frac{d P}{d \omega} =  2 \pi \int\limits_0^{\pi} \sin \theta d\theta
    \left(\frac{k^2}{d\omega/dk} P_k\right)_{k(\omega)},
\end{equation}
where $k(\omega)$ is the solution of $\omega=\omega_k^t$ and
$\theta$ is the polar angle of ${\bf k}$.

\section{Computation results}
Let us study the spectral and angular characteristics of second
harmonic plasma emission for various regimes of beam-plasma
interaction, which can be realized in mirror traps. We will vary the
plasma density from $2\cdot 10^{14}\, \mbox{cm}^{-3}$ to $5\cdot
10^{15}\, \mbox{cm}^{-3}$ for different fixed values of turbulence
energy $W/nT=0.01$, 0.05, 0.1 and for the fixed parameters $T=1$ keV
and $\Omega=0.2$.
\begin{figure*}[htb]
\bc\includegraphics[width=360bp]{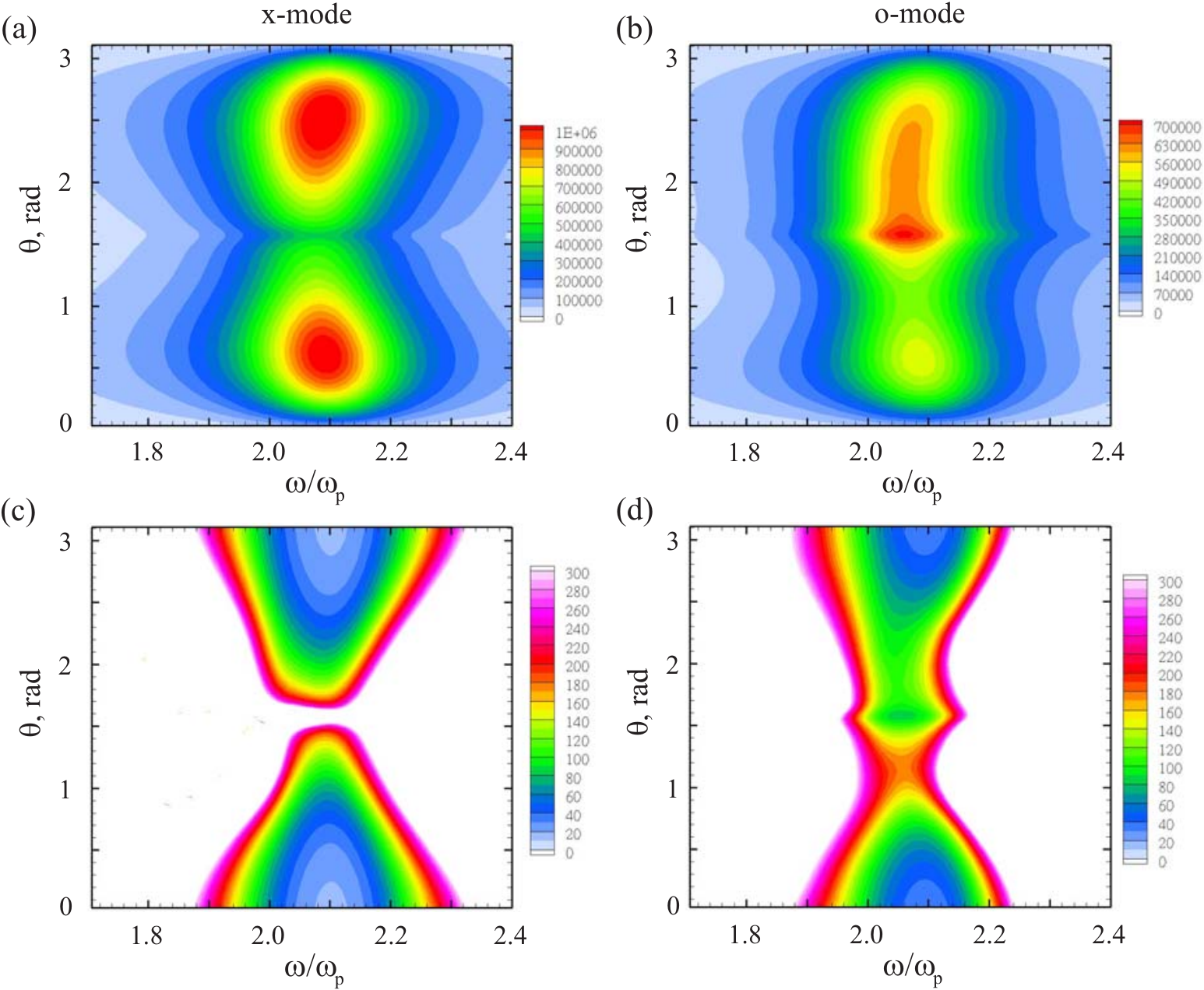} \ec \caption{ The
spectral power $dP/(d\omega d\theta)$ (in $\mbox{W/cm}^3$) for
x--mode emission (a) and $o$-mode emission (b). The path length (in
cm) for $x$-modes (c) and $o$-modes (d).}\label{r1}
\end{figure*}

We assume here that the source region of turbulent spectrum contains
the anisotropic population of resonant waves, which are directly
pumped by the beam, and the isotropic population of nonresonant
waves, which are the products of beam-driven waves scattering off
density fluctuations. In our computations, the isotropic part
occupies uniformly the spectral region $k c/\omega_p\in (0.1 ,k_m
c/\omega_p)$, where the upper bound corresponds to the typical
wavenumber of the modulation instability $k_m\simeq \sqrt{W/ n T}/
r_D$ and the lower bound excludes modes with wavelengths larger than
the typical plasma size. According to the model \cite{tim1}, the
anisotropic population of beam-driven waves contains a small
fraction of energy (10\%) and occupies the region: $kc/\omega_p \in
(1.1, 1.3)$ and $\theta \in (0,0.3)$.

In the regime with $n=3\cdot 10^{15}\, \mbox{cm}^{-3}$ and
$W/nT=0.05$, computation results for the spectral density of
emission power $dP/(d\omega d\theta)$ for extraordinary ($x$) and
ordinary ($o$) electromagnetic modes are presented in Figures
\ref{r1}(a) and \ref{r1}(b), respectively. It is seen that
$x$-mode emission is dominated by the contribution of obliquely
propagating waves with $\theta=30^0$ and $\theta=150^0$, whereas
the most intensive $o$-mode emission is concentrated near the
transverse to the magnetic field direction. Calculations of the
path length $l_k$ for these emissions are shown in Figures
\ref{r1}(c) and \ref{r1}(d). As one can see, the path length
reaches the minimal value 20-40 cm for longitudinally propagating
modes regardless of their polarization. For transverse
propagation, this length increases up to 90 cm for the $o$-mode
and exceeds 3 m for the $x$-mode. It means that emissions of both
electromagnetic modes, generated in the plasma column with the
diameter 5--6 cm typical to our beam-plasma experiments, are able
to escape from the plasma and can be used for different
applications.

Let us now find out how the frequency spectrum of electromagnetic
emission, described by $dP/d\omega$, depends on the plasma density.
\begin{figure}[htb]
\bc\includegraphics[width=220bp]{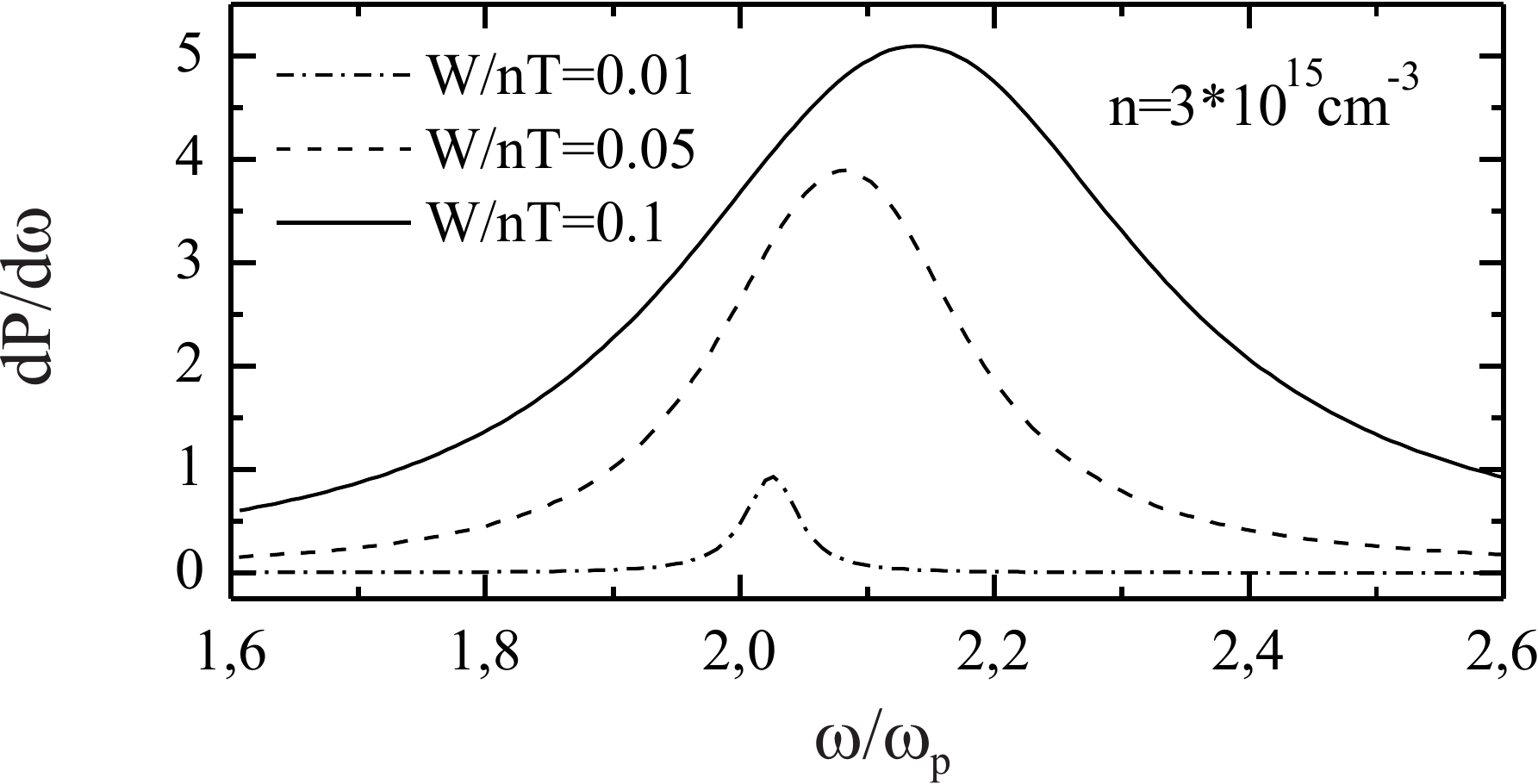} \ec \caption{The
spectral power of second harmonic emission $dP/d\omega$ in
$\mbox{MW/cm}^3$ for different values of turbulence energy
$W/nT$.}\label{r2}
\end{figure}
\begin{figure}[htb]
\bc\includegraphics[width=220bp]{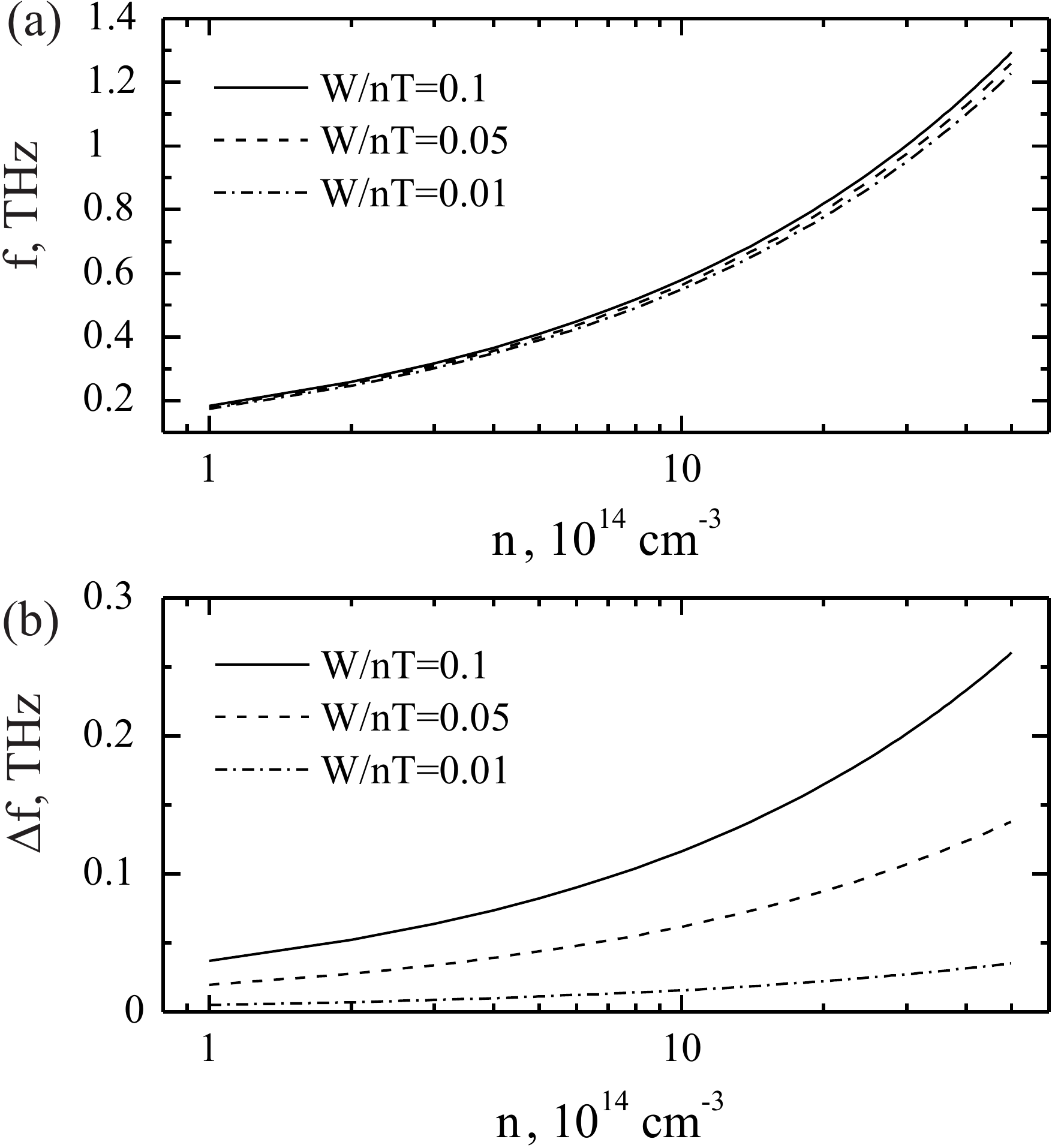} \ec \caption{
Dependences of $f$ (a) and $\Delta f$ (b) on the plasma density for
different values of turbulence energy $W/nT$.}\label{r3}
\end{figure}
As a function of dimensionless frequency $\omega/\omega_p$, the form
of this spectrum is affected by the level of turbulence only. For
the density $n=3\cdot 10^{15}\, \mbox{cm}^{-3}$ and different values
of turbulence energy, examples of frequency spectra, derived by
summation over polarizations, are shown in Figure \ref{r2}. In
dimensional form, the linear frequency $f$ corresponding to the
maximum of the spectral power $dP/d\omega$ and the width at half
maximum $\Delta f$ for a fixed turbulence energy depend on the
plasma density as follows: $f, \Delta f \propto n^{1/2}$. Figures
\ref{r3}(a) and \ref{r3}(b) demonstrate that $f$ and especially
$\Delta f$ grow with the increase of the turbulence level. It means
that the spectral width of plasma emission in regimes of rather
strong plasma turbulence is mainly determined by correlation
broadening of resonances in nonlinear three-wave interactions.

\begin{figure}[htb]
\bc\includegraphics[width=220bp]{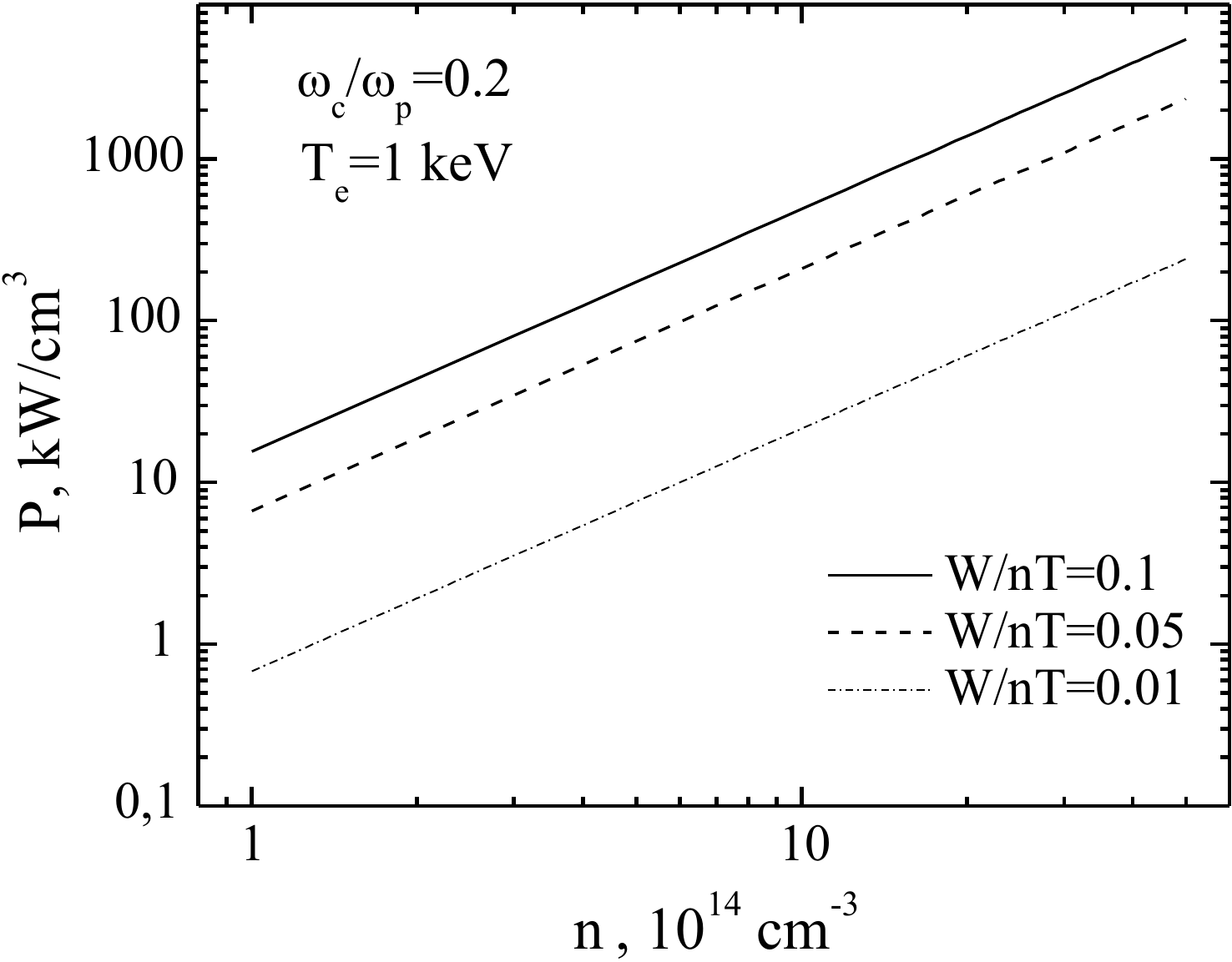} \ec \caption{ The
dependence of integral second harmonic emission power on the plasma
density for different values of turbulence energy $W/nT$.}\label{r4}
\end{figure}

To compute the integral power of second harmonic emission from the
unit volume of turbulent plasma, we integrate the function
$dP/d\omega$ over the frequency range $\omega/\omega_p \in
(1.6,2.6)$ and sum over polarizations. It is seen from Figure
\ref{r4},
that this power increases like $P\propto n^{3/2}$ with the
increase of plasma density for the fixed turbulence energy and
reaches the value 1 $\mbox{MW/cm}^3$ in the regime with $n=3\cdot
10^{15} \mbox{cm}^{-3}$ and $W/nT = 0.05$. It is interesting to
note that these calculations adequately describes recent
experimental results obtained in the strong magnetic field
$\Omega=0.8$ \cite{arz}. Indeed, the 1 $\mbox{kW/cm}^3$ range of
emission power, observed experimentally in the regime with
$n=2\cdot 10^{14} \mbox{cm}^{-3}$ and $W/nT = 0.01$, is also
reproduced by calculations presented in Figure \ref{r4}. From
comparison with the results of \cite{tim1} we can conclude that
the external magnetic field broadens the spectral width of
electromagnetic emission, but does not change drastically the
integral power. Since the pumping power (\ref{eq1}) has the same
dependence on plasma density as the emission power, the conversion
efficiency $\epsilon$ in our model is completely determined by the
turbulence energy. For $W/nT=0.05$, this efficiency is estimated
as $\epsilon\simeq 1\%$.

\begin{figure}[htb]
\bc\includegraphics[width=220bp]{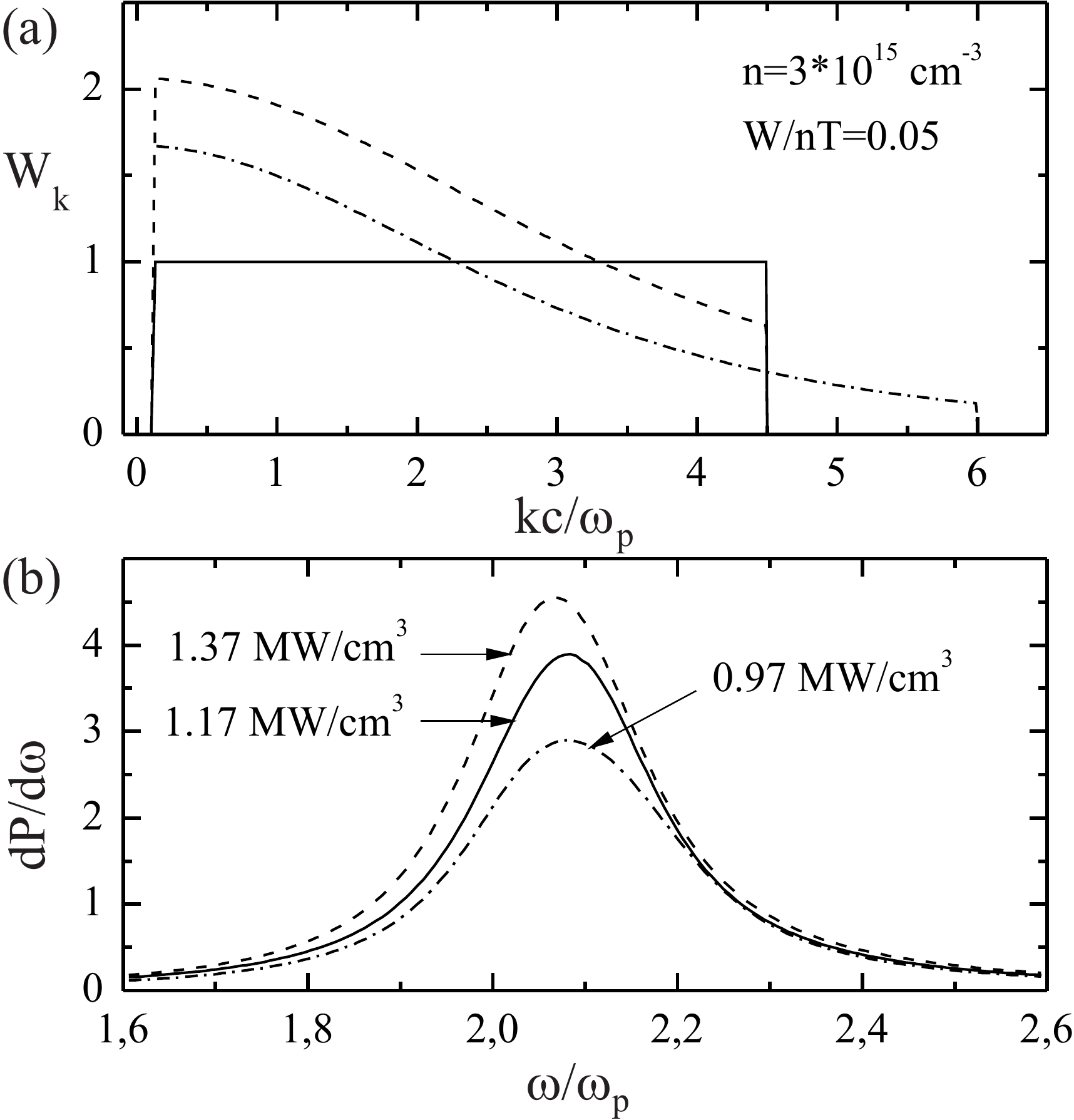} \ec
\caption{Different shapes of the isotropic part of turbulent
$k$-spectra (a) and the corresponding spectral power of
electromagnetic emission (b).}\label{r5}
\end{figure}

To make clearer understanding how computation results are
sensitive to the shape of turbulent spectrum, we calculate the
spectral emission power for different spectra corresponding to the
same turbulence energy (Figure \ref{r5}). When we replace the
uniform isotropic spectrum of upper-hybrid modes used in previous
calculations on more realistic falling spectra, changes in the
integral emission power do not exceed 20\%.

It should be emphasized that the high value of turbulence energy
$W/nT=0.05$, which is required to generate 1 THz emission with the
power density 1 $\mbox{MW/cm}^3$, can be achieved in beam-plasma
experiments at the GOL-3 multimirror trap. Indeed, for the typical
beam energy 1 MeV and current density 1.5 $\mbox{kA/cm}^2$, the
pumping power reaches the required level of 100 $\mbox{MW/cm}^3$
inside the region of most intensive beam-plasma interaction, where
large amplitude coherent wave-packets are excited. Since the length
of this region is estimated as $l\sim v_b/\Gamma\simeq 1-3$ cm, the
total power of terahertz emission in our laboratory experiments can
reach 30--100 MW.

\section{Conclusion}
The spectral power of second harmonic electromagnetic emission,
generated in the long-wavelength region of strong plasma turbulence
due to coalescence of upper-hybrid waves, was calculated for
different regimes of beam-plasma interaction, which can be realized
in modern experiments at mirror traps. Computation results have
shown that beam energy transferred to plasma oscillations in the
regime of strong plasma turbulence can be converted efficiently to
electromagnetic radiation. We have studied spectral characteristics
of plasma emission in wide ranges of plasma density and turbulence
energy and shown that the power of terahertz radiation in our
beam-plasma experiments can reach tens of MW. Taking into account
the angular distribution of this emission and possibility to change
the emission frequency by varying the plasma density we come to the
conclusion that terahertz radiation generated in a beam-driven
strongly turbulent plasma can be very attractive for different
applications.

 \ack Authors thank I.A.
Kotelnikov for useful discussions. This work is supported by grant
11.G34.31.0033 of the Russian Federation Government, President
grant NSh-5118.2012.2, Russian Ministry of Education and Science
and RFBR grants 11-02-00563, 11-01-00249.

\end{document}